\documentclass[final,3p,twocolumn,times]{elsarticle}%
\usepackage{amssymb}
\usepackage{amsmath}
\usepackage{amsfonts}
\usepackage[usenames]{color}
\usepackage{graphicx}%
\providecommand{\U}[1]{\protect\rule{.1in}{.1in}}
\journal{Solid State Communications}
\begin{document}
\begin{frontmatter}
\title{Feynman's path-integral polaron treatment approached using time-ordered operator calculus$^*$}
\author[1,2]{S. N. Klimin}
\author[1,3]{J. T. Devreese}
\ead{jozef.devreese@ua.ac.be}
\address[1]{Theorie van Kwantumsystemen en Complexe Systemen (TQC),
Universiteit Antwerpen, Groenenborgerlaan 171, B-2020 Antwerpen, Belgium}
\address[2]{Department of Theoretical Physics, State University of
Moldova, MD-2009 Chisinau, Moldova}
\address[3]{COBRA, Eindhoven University of Technology, 5600 MB Eindhoven,
The Netherlands}
\begin{abstract}
The Feynman all-coupling variational approach for the polaron is re-formulated and extended
using the Hamiltonian formalism with time-ordered operator calculus. Special attention is devoted
to the excited polaron states. The energy levels and the inverse lifetimes of the excited polaron states
are, for the first time, explicitly derived within this all-coupling approach.
Remarkable agreement of the obtained transition energies with the peak positions of the polaron
optical conductivity calculated using diagrammatic quantum Monte Carlo is obtained.
\end{abstract}
\end{frontmatter}

Several studies have been devoted to the search of a Hamiltonian formalism
equivalent to Feynman's path integral approximation to polaron theory.
Bogolubov \cite{Bogolubov} reproduced the Feynman result for the polaron free
energy \cite{Feynman} using time-ordering T-products . Yamazaki
\cite{Yamazaki} introduced two kinds of auxiliary vector fields to derive
Feynman's ground state polaron energy expression with the operator technique,
however he found no proof of the variational nature of this result. Cataudella
\emph{et al}. \cite{Cataudella} formally re-obtained Feynman's polaron
ground-state energy expression by introducing additional degrees of freedom,
but again their result could not be proved to constitute an upper bound for
the polaron ground state energy.

The study of the excited polaron states is of interest i.~a. for its
application to the polaron response properties. In \cite{FHIP} a path-integral
based response-formalism was introduced that was applied to derive polaron
optical absorption spectra in \cite{Devreese1972}. The results for the polaron
response obtained in \cite{Devreese1972} were re-derived with a Hamiltonian
technique (Mori- formalism) in \cite{PD1983}.

To the best of our knowledge, no explicit description of the polaron excited
states has been derived within the \textquotedblleft all
coupling-\textquotedblleft\ Feynman approach. Only for the limiting cases of
weak and strong coupling approximations (and for a 1D-model system) such
excitation spectra were derived \cite{DE1,DE2}.

In principle, the spectrum of the polaron excited states can be derived
indirectly -- using a Laplace transform of the finite-temperature partition
function. However, it is not clear how to realize this program in practice.

The polaron excitation spectrum is interesting by itself. E.g. the existence
and the nature of \textquotedblleft relaxed excited states\textquotedblright,
\textquotedblleft Franck-Condon states\textquotedblright, \textquotedblleft
scattering states\textquotedblright\ is understood from the mathematical
structure of corresponding eigenstates.

In the present letter we first present a re-derivation of the original Feynman
variational path integral polaron model \cite{Feynman} for the ground state,
using a Hamiltonian formalism, and we do provide a proof of the upper bound
nature of the obtained ground state energy. Furthermore, using Feynman's
(Hamiltonian-) time-ordered operator calculus (and an \emph{ad hoc} unitary
transformation) we obtain explicitly -- and for the first time -- the excited
polaron states that correspond to the Feynman polaron model.

The novelty of the present approach consists (a) in the \emph{direct
calculation of the energies and the lifetimes of the excited polaron states}
(within a Hamiltonian all-coupling approach -- developed in this work --
equivalent to the Feynman path integral polaron model) and (b) in the
extension of the Feynman variational technique to non-parabolic trial
potentials. Although the time-ordered operator calculus is formally equivalent
to the path-integral formalism, it is not obvious how to directly calculate
the excited polaron states using path integrals.

The present work, formulated with the (Hamiltonian) time ordered operator
calculus, thus provides an (equivalent) tool complementary with respect to the
Feynman path integral approach to the polaron, to study the polaron problem.
Additionally we directly study the excited polaron states.

Consider an electron-phonon system with the Fr\"{o}hlich Hamiltonian%

\begin{align}
H &  =\frac{\mathbf{p}^{2}}{2}+H_{ph}+H_{e-ph},\label{H}\\
H_{ph} &  =\sum_{\mathbf{q}}\left(  a_{\mathbf{q}}^{+}a_{\mathbf{q}}+\frac
{1}{2}\right)  ,\label{Hp}\\
H_{e-ph} &  =\frac{1}{\sqrt{V}}\sum_{\mathbf{q}}\frac{\sqrt{2\sqrt{2}\pi
\alpha}}{q}\left(  a_{\mathbf{q}}+a_{-\mathbf{q}}^{+}\right)
e^{i\mathbf{q\cdot r}}.\label{Hep}%
\end{align}
Here, the Feynman units are used: $\hbar=1,$ the band mass $m_{b}=1,$ the
LO-phonon frequency $\omega_{\mathrm{LO}}=1$.

The polaron partition function after exact averaging over phonon states is%
\begin{equation}
Z_{pol}=\mathtt{Tr}\left[  \mathtt{T}\exp\left(  -\int_{0}^{\beta}%
\frac{\mathbf{p}_{\tau}^{2}}{2}d\tau+\hat{\Phi}\left[  \mathbf{r}_{\tau
}\right]  \right)  \right]  , \label{zpol}%
\end{equation}
where $\beta=\frac{1}{k_{B}T}$. The \textquotedblleft influence
phase\textquotedblright\ of the phonons $\hat{\Phi}\left[  \mathbf{r}_{\tau
}\right]  $ in the the time-ordered operator calculus has the same form as in
the path-integral representation. The polaron free energy is determined as%
\begin{equation}
F_{pol}=-\frac{1}{\beta}\ln Z_{pol}. \label{fpol}%
\end{equation}

The trial Hamiltonian describes the electron interacting with a fictitious
particle of the mass $m_{f}$ through an attractive potential $V_{f}$:
\begin{equation}
H_{tr}=\frac{\mathbf{p}^{2}}{2}+\frac{\mathbf{p}_{f}^{2}}{2m_{f}}+V_{f}\left(
\mathbf{r}-\mathbf{r}_{f}\right)  . \label{Htr1}%
\end{equation}
The trial potential $V_{f}$ is, in general, non-parabolic. The parabolic
potential with frequency parameter $w$ corresponds to the Feynman polaron model.

Consider the \textquotedblleft extended\textquotedblright\ partition function
of the electron-phonon system%
\begin{equation}
Z_{ext}=Z_{f}Z_{pol} \label{aa}%
\end{equation}
where $Z_{f}$ is the partition function of a fictitious particle,%
\begin{equation}
Z_{f}\equiv\mathtt{Tr}\left[  \mathtt{T}\exp\left(  -\int_{0}^{\beta}d\tau
H_{f,\tau}\right)  \right]  , \label{zf}%
\end{equation}
with Hamiltonian%
\begin{equation}
H_{f}=\frac{\mathbf{p}_{f}^{2}}{2m_{f}}+V_{f}\left(  \mathbf{r}_{f}\right)  .
\label{Hf}%
\end{equation}

The polaron free energy is expressed as the difference%
\begin{equation}
F_{pol}=F_{ext}-F_{f}, \label{fp}%
\end{equation}
where $F_{f}$ is the free energy of the fictitious particle confined to the
potential $V\left(  \mathbf{r}_{f}\right)  $. The free energies $F_{ext}$ and
$F_{f}$ are determined similarly to (\ref{fpol}), with corresponding partition
functions. In the zero-temperature limit, the free energies $F_{pol}$,
$F_{ext}$ and $F_{f}$ become, respectively, the ground-state energies
$E_{pol}^{0}$, $E_{ext}^{0}$ and $E_{f}^{0}$.

The key element of the present approach is the unitary transformation%
\begin{equation}
U=e^{-i\mathbf{p}_{f}\cdot\mathbf{r}}. \label{U}%
\end{equation}
Application of this canonical transformation results in the transformed
\textquotedblleft extended\textquotedblright\ Hamiltonian $H_{ext}^{\prime
}=UH_{ext}U^{-1}$,%
\begin{align}
H_{ext}^{\prime}  &  =\frac{\left(  \mathbf{p}+\mathbf{p}_{f}\right)  ^{2}}%
{2}+\frac{\mathbf{p}_{f}^{2}}{2m_{f}}+V_{f}\left(  \mathbf{r}_{f}%
-\mathbf{r}\right) \nonumber\\
&  +H_{ph}+H_{e-ph}. \label{Hext}%
\end{align}
This Hamiltonian can be represented as a sum of an unperturbed Hamiltonian
\begin{equation}
H_{0}\equiv H_{tr}+H_{ph} \label{H0}%
\end{equation}
and an interaction term%
\begin{equation}
V\equiv\frac{1}{2}\mathbf{p}_{f}^{2}+\mathbf{p}\cdot\mathbf{p}_{f}+H_{e-ph}.
\label{V}%
\end{equation}

Further we use the variational principle for the ground-state energy in terms
of the time-ordered operators following Ref. \cite{DB1992}. The exact ground
state $\left\vert 0\right\rangle $ of the system with the Hamiltonian
(\ref{Hext}) can be written in the interaction representation starting from
the unperturbed ground state $\left\vert -\infty\right\rangle $:%
\begin{equation}
\left\vert 0\right\rangle =\mathcal{U}\left(  \infty,-\infty\right)
\left\vert -\infty\right\rangle \label{0}%
\end{equation}
where $\mathcal{U}\left(  \infty,-\infty\right)  $ is the time-evolution
operator,%
\begin{equation}
\mathcal{U}\left(  t_{2},t_{1}\right)  =\mathcal{T}\exp\left(  -i\int_{t_{1}%
}^{t_{2}}e^{-\delta\left\vert t\right\vert }e^{iH_{0}t}Ve^{-iH_{0}t}\right)  .
\label{T}%
\end{equation}
Here, $\delta\rightarrow+0$ and $\mathcal{T}$ denotes time ordering.

In the exact expectation value for the ground state energy $E_{ext}^{0}%
\equiv\left\langle 0\left\vert H_{ext}^{\prime}\right\vert 0\right\rangle $,
the phonons are eliminated using the time ordered-operator calculus as in Ref.
\cite{DB1992}. The average of the interaction term becomes then%
\begin{align}
&  \left\langle 0\left\vert H_{e-ph}\right\vert 0\right\rangle \nonumber\\
&  =-i\frac{\sqrt{2}\pi\alpha}{V}\int_{-\infty}^{\infty}dte^{-i\left\vert
t\right\vert -\delta\left\vert t\right\vert }\nonumber\\
&  \times\sum_{\mathbf{q}}\frac{1}{q^{2}}\left\langle \infty\left\vert
\mathcal{T}\left[  \mathcal{U}\left(  \infty,-\infty\right)  e^{i\mathbf{q}%
\cdot\left[  \mathbf{r}\left(  t\right)  -\mathbf{r}\left(  0\right)  \right]
}\right]  \right\vert -\infty\right\rangle .\label{Av}%
\end{align}
This means that the polaron ground state energy is exactly described using a
retarded potential in the interaction representation, cf. Eq. (2.16) of Ref.
\cite{DB1992}.

The ground state energy satisfies the Ritz variational principle with a trial
state. Choosing the trial state as the ground state of the Hamiltonian
(\ref{H0}), the variational principle can be written as \cite{DB1992}%
\begin{align}
E_{ext}^{0}  &  \leq E_{tr}^{0}\nonumber\\
&  +\left\langle \infty\left\vert \mathcal{T}\left\{  \mathcal{U}_{tr}\left(
\infty,-\infty\right)  \left[  H_{ext}^{\prime}\left(  0\right)  -H_{0}\left(
0\right)  \right]  \right\}  \right\vert -\infty\right\rangle , \label{FVP}%
\end{align}
where $\mathcal{U}_{tr}\left(  \infty,-\infty\right)  $ is the time-evolution
operator corresponding to the trial Hamiltonian (\ref{Htr1}).

The exact polaron ground state energy is denoted here as $E^{0}\left(
\mathbf{k}\right)  $, where $\mathbf{k}$ is the polaron translation momentum.
We find an upper bound for $E^{0}\left(  \mathbf{k}\right)  $ substituting
(\ref{Av}) in (\ref{FVP}) and using the exact wave functions and energy levels
of the trial Hamiltonian. The trial Hamiltonian (\ref{Htr1}) can be rewritten
in terms of the coordinates $\left(  \mathbf{R},\boldsymbol{\rho}\right)  $
and momenta $\left(  \mathbf{P},\vec{\pi}\right)  $ of the center-of-mass and
relative (internal) motions of the trial system with the masses $M=1+m_{f}$
and $\mu=m_{f}/\left(  1+m_{f}\right)  $ using the frequency $v=wM$. The
energy spectrum of the trial system is the sum of the translation- and
oscillation contributions,%
\begin{equation}
E_{\mathbf{k},n}=\frac{\mathbf{k}^{2}}{2M}+\varepsilon_{n},\;\varepsilon
_{n}=v\left(  n+\frac{3}{2}\right)  .\label{ener}%
\end{equation}
The eigenfunctions of the Hamiltonian (\ref{Htr1}) are products of
translational- and oscillatory wave functions:%
\begin{equation}
\psi_{\mathbf{k};l,n,m}\left(  \mathbf{R},\boldsymbol{\rho}\right)  =\frac
{1}{\sqrt{V}}e^{i\mathbf{k\cdot R}}\varphi_{l,n,m}\left(  \boldsymbol{\rho
}\right)  ,\label{psi}%
\end{equation}
where $\varphi_{l,n,m}\left(  \boldsymbol{\rho}\right)  $ is the 3D
harmonic-oscillator wave function with a given angular momentum. The result is%
\begin{align}
E^{0}\left(  \mathbf{k}\right)   &  \leq\mathcal{E}_{p}^{\left(  0,0\right)
}\left(  \mathbf{k}\right)  ,\\
\mathcal{E}_{p}^{\left(  0,0\right)  }\left(  \mathbf{k}\right)   &  =\frac
{3}{4}\frac{\left(  v-w\right)  ^{2}}{v}\nonumber\\
&  +\frac{1}{2}\left(  1-\frac{1}{\left(  1+m_{f}\right)  ^{2}}\right)
\mathbf{k}^{2}-\frac{\sqrt{2}\alpha}{4\pi^{2}}\int\frac{d\mathbf{q}}{q^{2}%
}\nonumber\\
&  \times\sum_{\mathbf{k}^{\prime},l^{\prime},n^{\prime},m^{\prime}}%
\frac{\left\vert \left\langle \psi_{\mathbf{k};0,0,0}\left\vert
e^{i\mathbf{q\cdot r}}\right\vert \psi_{\mathbf{k}^{\prime};l^{\prime
},n^{\prime},m^{\prime}}\right\rangle \right\vert ^{2}}{\frac{1}{2\left(
m_{f}+1\right)  }\left(  \left(  \mathbf{k}^{\prime}\right)  ^{2}%
-\mathbf{k}^{2}\right)  +vn^{\prime}+1},\label{E0}%
\end{align}
where $v>w$ are the Feynman variational frequencies. The functional (\ref{E0})
can be reduced to the known Feynman result for the polaron ground-state
energy. In the r.h.s. of (\ref{E0} at the polaron momentum $\mathbf{k}=0$, we
introduce the integral over the Euclidean time:
\begin{equation}
\frac{1}{\frac{\left(  \mathbf{k}^{\prime}\right)  ^{2}}{2\left(
m_{f}+1\right)  }+vn^{\prime}+1}=\int_{0}^{\infty}e^{-\left(  \frac{\left(
\mathbf{k}^{\prime}\right)  ^{2}}{2\left(  m_{f}+1\right)  }+vn^{\prime
}+1\right)  \tau}d\tau.\label{Integr}%
\end{equation}
After this, the summations and integrations in (\ref{Epol}) are performed
analytically, and we arrive at the Feynman variational expression for the
polaron ground-state energy:%
\begin{align}
\left.  E^{0}\left(  \mathbf{k}\right)  \right\vert _{\mathbf{k}=0} &
\leq\frac{3}{4}\frac{\left(  v-w\right)  ^{2}}{v}\nonumber\\
&  -\frac{\alpha v}{\sqrt{\pi}}\int_{0}^{\infty}\frac{e^{-\tau}}{\sqrt
{w^{2}\tau+\frac{v^{2}-w^{2}}{v}\left(  1-e^{-v\tau}\right)  }}d\tau
.\label{EF}%
\end{align}

The electron-phonon contribution in (\ref{E0}) is structurally similar to the
second-order perturbation correction to the polaron ground-state energy due to
the electron-phonon interaction (using states of the Feynman model
$\psi_{\mathbf{k};l,n,m}$ as the zero-order approximation). Therefore we can
estimate the energies of the excited polaron states when averaging the
difference between exact and unperturbed Hamiltonians in (\ref{FVP}) with an
excited trial state. We then arrive at the following extension for the r.h.s.
of (\ref{E0}):%
\begin{align}
\mathcal{E}_{p}^{\left(  l,n\right)  }\left(  \mathbf{k}\right)   &
=\frac{v^{2}+w^{2}}{2v}\left(  n+\frac{3}{2}\right)  -\frac{3}{2}w\nonumber\\
&  +\frac{1}{2}\left(  1-\frac{1}{\left(  1+m_{f}\right)  ^{2}}\right)
\mathbf{k}^{2}-\frac{\sqrt{2}\alpha}{4\pi^{2}}\int\frac{d\mathbf{q}}{q^{2}%
}\nonumber\\
&  \times\sum_{\mathbf{k}^{\prime},l^{\prime},n^{\prime},m^{\prime}}%
\frac{\left\vert \left\langle \psi_{\mathbf{k};l,n,m}\left\vert
e^{i\mathbf{q\cdot r}}\right\vert \psi_{\mathbf{k}^{\prime};l^{\prime
},n^{\prime},m^{\prime}}\right\rangle \right\vert ^{2}}{\frac{1}{2\left(
m_{f}+1\right)  }\left(  \left(  \mathbf{k}^{\prime}\right)  ^{2}%
-\mathbf{k}^{2}\right)  +v\left(  n^{\prime}-n\right)  +1}. \label{Epol}%
\end{align}

In the same approach, we obtain the inverse lifetimes for the excited states
of the polaron:%
\begin{align}
\Gamma_{l,n}\left(  \mathbf{k}\right)   &  =\frac{\sqrt{2}\alpha}{4\pi}%
\sum_{\mathbf{k}^{\prime},l^{\prime},n^{\prime},m^{\prime}}\int d\mathbf{q}%
\frac{1}{q^{2}}\nonumber\\
&  \times\left\vert \left\langle \psi_{\mathbf{k};l,n,m}\left\vert
e^{i\mathbf{q\cdot r}}\right\vert \psi_{\mathbf{k}^{\prime};l^{\prime
},n^{\prime},m^{\prime}}\right\rangle \right\vert ^{2}\nonumber\\
&  \times\delta\left(  \frac{q^{2}}{2\left(  m_{f}+1\right)  }+v\left(
n^{\prime}-n\right)  +1\right)  . \label{gamma}%
\end{align}
The broadening of the excited polaron \textquotedblleft
non-scattering\textquotedblright\ states must be taken into account for an
analytical study of the polaron optical conductivity.

Using the above expressions, we determine the transition energies for the
transitions between the ground and the first excited state $\hbar
\Omega_{0\rightarrow1exc}\equiv E_{p}^{\left(  1exc\right)  }-E_{p}^{\left(
0\right)  }$. Let us first consider the transition energies in which
$E_{p}^{\left(  1exc\right)  }$ are calculated using optimal values of the
parameters of the Feynman model obtained from the minimization of the
variational ground-state energy $E_{p}^{\left(  0\right)  }$. This method
formally leads to the Franck-Condon (FC) excited states, with the
\textquotedblleft frozen\textquotedblright\ phonon configuration corresponding
to the ground state of the polaron. Note that the existence of Franck-Condon
states as eigenstates of the Fr\"{o}hlich polaron Hamiltonian has not been
proved: Ref \cite{DE2} suggests their non-existence as eigenstates for a
simplified polaron model. Nevertheless the Franck-Condon concept can be
significant, e.~g. for approximate treatments using a basis of Franck-Condon
states, as indicative for the frequency of the maxima of phonon-sidebands, etc.%

\begin{figure}
[h]
\begin{center}
\includegraphics[
height=5.7903cm,
width=6.9589cm
]%
{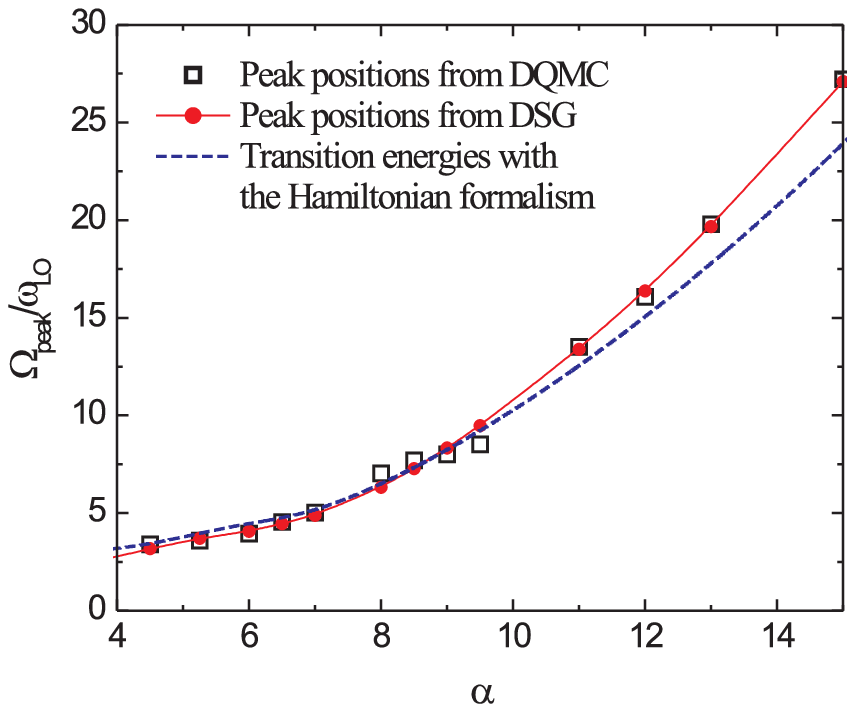}%
\caption{Franck-Condon transition energies as a function of the coupling
constant compared to the lowest-energy peak position of the polaron optical
conductivity from Ref. \cite{Devreese1972} and the maximum of the polaron
optical conductivity band from Ref. \cite{Mishchenko2003}.}%
\end{center}
\end{figure}

In Fig. 1, the FC transition energies calculated with the approach introduced
in the present work for polaron momentum $\mathbf{k}=0$ are plotted as a
function of the coupling constant $\alpha$. They are compared with the peak
energies of the polaron optical conductivity calculated using the diagrammatic
Monte Carlo method (DQMC) \cite{Mishchenko2003,DeFilippis2006} and with the
peak energies attributed to polaron \textquotedblleft relaxed excited
states\textquotedblright\ (RES) in Ref. \cite{Devreese1972} (\textquotedblleft
DSG\textquotedblright). The DQMC and DSG main-peak energies are close to each
other in the whole range of the coupling strength. In the range $4\lessapprox
\alpha\lessapprox10$, the present result for the transition energy is close to
the DQMC and the DSG peak energies. Furthermore, in this range of $\alpha$,
the non-monotonous behavior of the curvature is remarkably the same for the
DQMC and DSG peak energies and for the present result.

There is a remarkable agreement between the peaks attributed to the RES in
Ref. \cite{Devreese1972}, the peak positions obtained within the
strong-coupling approach, Eq. (3) of Ref. \cite{DeFilippis2006}, and the
positions of the maximum of the optical conductivity band calculated in Ref.
\cite{Mishchenko2003} using DQMC. It is reasonable that the three aforesaid
peaks must be interpreted in one and the same way. In order to clarify this,
we can refer to Ref. \cite{Mishchenko2003}. In the strong-coupling regime, the
dominant broad peak of the polaron optical conductivity spectrum can be
considered as a \textquotedblleft Franck-Condon sideband\textquotedblright\ of
the \textquotedblleft groundstate to RES-transition\textquotedblright, even if
this latter transition can have a negligible oscillator strength (see also
\cite{KED1969}). The optical conductivity spectra of Ref.
\cite{DeFilippis2006} in the strong-coupling approximation have been
calculated taking into account the polaronic shift of the energy levels. The
polaronic shift in Ref. \cite{DeFilippis2006} has been calculated with the
Franck-Condon wave functions (i. e., with the strong-coupling wave functions
corresponding to the \textquotedblleft frozen\textquotedblright\ lattice
configuration for the ground state). Note that the exact excitation spectrum
of the Fr\"{o}hlich-Hamiltonian might be devoid of Franck-Condon eigenstates,
cf. Ref. \cite{DE2}). It should be remarked that the maxima of the FC-sideband
structures of Ref. \cite{Devreese1972} are positioned at the frequency
$\Omega=v$, i. e., at the transition frequency for the model system without
the polaron shift.

The Franck-Condon peak energies calculated in the present work also take into
account the polaron shift. As follows from the above analysis, in the
strong-coupling limit they must correspond to the Franck-Condon peak energies
of the strong-coupling expansion of Ref. \cite{DeFilippis2006}. The agreement
of the position of the maxima of these peaks with those attributed to
transitions to the RES in Ref. \cite{Devreese1972} shows that in the
strong-coupling range of $\alpha$, the latter should be associated to the
Franck-Condon sidebands rather than to the RES.

Another approach, in which the parameters of the first excited state are
determined self-consistently (Ref. \cite{KED1969}), was used i.~a. to
calculate (in the strong-coupling case) the (lowest) energy level of the
relaxed excited state (RES). The transitions from the polaron ground state to
the RES correspond to a zero-phonon peak in the optical conductivity.

For the study of the energies of excited states of the polaron, a variational
approach requires special care, because the excited states of the polaron are
not stable. A variational approach, strictly speaking, is only valid for
excited states when the variational wave function of the excited state is
orthogonal to the exact ground-state wave function.

For the estimation of the energy of the first RES with our present formalism,
we determine a minimum of the expression (\ref{Epol}) in a physically
reasonable range of the variational parameters. In order to determine that
range, we refer to Ref. \cite{Lepine}, where the energy of the polaron RES is
calculated variationally within the Green's function formalism.

The expression for the RES energy in Ref. \cite{Lepine} contains the
electron-phonon contribution corresponding to the second-order perturbation
formula. It differs, however, from the weak-coupling second-order perturbation
expression by the choice of the unperturbed states: in Ref.\cite{Lepine} those
are variational states rather than free-electron states. There exists some
analogy between our approach and that of Ref. \cite{Lepine}. The latter,
however, does not take into account the translation invariance of the polaron problem.

In Ref. \cite{Lepine}, the energy of the polaron RES is calculated
variationally. The unperturbed wave function of the RES is chosen orthogonal
(due to symmetry) to the unperturbed ground state wave function. In the
present approach, this orthogonality is also exactly satisfied because of symmetry.

The expressions for the polaron RES energy of Ref. \cite{Lepine} contain
singularities, which occur when the energies of the unperturbed ground state
and that of the first excited states are in resonant with the LO-phonon
energy. These singularities are related to the instability of the excited
polaron with respect to the emission of LO-phonons. Using the same reasoning
as in Ref. \cite{Lepine} we search for a local minimum of the polaron RES
energy in the range where the confinement frequency $v$ of the Feynman model
satisfies the inequality $v>1$. The instability of the excited polaron state
is then avoided.

The resulting numerical values of the transition energy to the first RES as a
function of $\alpha$ are shown in Fig. 2. They are compared with the
numerical-DQMC peak energies of the polaron optical conductivity band
\cite{Mishchenko2003,DeFilippis2006}, with the FC transition energies obtained
in the present work, and with the leading term of the strong-coupling
approximation for the RES transition energy from Ref. \cite{KED1969}.%

\begin{figure}
[h]
\begin{center}
\includegraphics[
height=5.7903cm,
width=6.9589cm
]%
{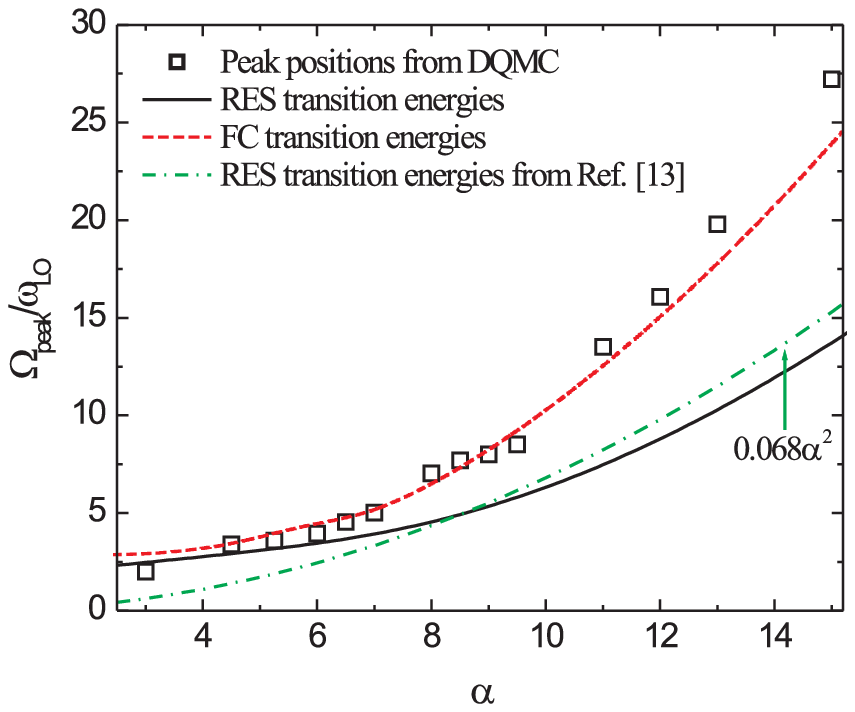}%
\caption{The transition energy for the transition from the polaron ground
state to the first RES(\emph{solid black curve}) and to the first excited FC
state (\emph{dashed red curve}) as a function of $\alpha$ obtained in the
present work, compared with the maximum of the polaron optical conductivity
band from numerical DQMC (\emph{black squares}, Ref. \cite{Mishchenko2003}).
\emph{The dashed-dot green curve}: the strong-coupling result for this
transition energy as given in Ref. \cite{KED1969}.}%
\end{center}
\end{figure}

For $\alpha\lesssim2.5$, there exists no minimum of $E_{p}^{\left(
1exc\right)  }$ in the range $v>1$. We can interpret this result as a
manifestation of the fact that for decreasing coupling strength, the RES is
suppressed at sufficiently weak coupling. We see that for sufficiently small
$\alpha$ ($\alpha\lesssim6$), the RES transition energies show good agreement
with the DQMC peak energies, what confirms the concept of RES developed in
Refs. \cite{Devreese1972,KED1969}. For higher coupling strengths, the DQMC
data appear to be closer to the FC (rather than to RES) transition energies.
This result can be an indication of the fact that with increasing $\alpha$,
the mechanism of the polaron optical absorption changes its nature as
suggested in Ref. \cite{DeFilippis2006}, from a regime with dynamic lattice
relaxation (for which the RES are relevant) at weak and intermediate coupling
to the Franck-Condon (\textquotedblleft LO-phonon sidebands\textquotedblright%
-) regime at strong coupling.

In summary, we have re-formulated the Feynman all-coupling path integral
method for the polaron problem within a Hamiltonian formalism using
time-ordered operator calculus. This reformulation allows us to describe not
only the free energy and the ground state, but also to directly determine --
for the first time -- the excited polaron states that correspond to the
Feynman all-coupling polaron model. A variational procedure for the polaron
RES energy has been developed, within the formalism presented in this work,
which provides results i.a. in agreement with the strong-coupling limit of
Ref. \cite{KED1969}. The present treatment offers the prospect of further
elucidation of the nature of the polaron resonances (\textquotedblleft relaxed
excited states\textquotedblright\ versus \textquotedblleft Franck-Condon
sidebands\textquotedblright\ [9]) at intermediate coupling.

This work was supported by FWO-V projects G.0356.06, G.0370.09N, G.0180.09N,
G.0365.08, and the WOG WO.035.04N (Belgium).

\end{document}